\documentclass[aps, prd, a4paper, floats, floatfix, nofootinbib, showpacs, twocolumn, amsfonts, amssymb, amsmath]{revtex4-1}
\usepackage{amssymb,latexsym}
\usepackage{amsmath, amsthm}

\usepackage{amscd}
\usepackage{times}
\usepackage{epsfig}
\usepackage{psfrag}
\usepackage{graphicx}
\usepackage{overpic}

\begin{document}

\title{Non-singular quantum-inspired gravitational collapse}

\author{Cosimo Bambi}
\email{bambi@fudan.edu.cn}

\author{Daniele Malafarina}
\email{daniele@fudan.edu.cn}

\author{Leonardo Modesto}
\email{lmodesto@fudan.edu.cn}

\affiliation{Center for Field Theory and Particle Physics \& Department of Physics, 
Fudan University, 200433 Shanghai, China}

\date{\today}

\swapnumbers

\begin{abstract}
We consider general relativistic homogeneous gravitational collapses for dust 
and radiation. We show that replacing the density profile with an effective density 
justified by some quantum gravity framework leads to the avoidance of the final 
singularity. The effective density acts on the collapsing cloud by introducing an 
isotropic pressure, which is negligible at the beginning of the collapse and 
becomes negative and dominant in the strong field regime. Event horizons
never form and therefore the outcome of the collapse is not a black hole, in the
sense that there are no regions causally disconnected from future null infinity.
Apparent horizons form when the mass of the object exceeds a critical value,
disappear when the matter density approaches an upper bound and gravity 
becomes very weak (asymptotic freedom regime), form again after the bounce 
as a consequence of the decrease in the matter density, and eventually 
disappear when the density becomes too low and the matter is radiated away. 
The possibility of detecting radiation coming from the high density region of 
a collapsing astrophysical object in which classically there would be the 
creation of a singularity could open a new window to experimentally test 
theories of quantum gravity.
\end{abstract}


\maketitle

\section{Introduction}

The search for a theory of quantum gravity is surely one of the most important
open issues in contemporary theoretical high energy physics and a very active
research field. The key problem is the complete absence of experimental data 
capable of testing the validity of the large number of different models that have been proposed 
so far. Up to now, we have no observational evidence of any quantum gravity 
phenomenon. The natural energy scale of quantum gravity is the Planck mass 
$M_{\rm Pl} \sim 10^{19}$~GeV, which is definitively too high to be reached in 
particle colliders on Earth, even in the foreseeable future.

In the literature, there are a few proposals that try to catch observational signatures 
of quantum gravity effects. The most promising approach is likely the study of some 
primordial features in the cosmic microwave background radiation; they are supposed 
to have been generated during inflation and may encode some details about quantum
gravity~\cite{lqg1}. Another proposal concerns the possibility of the existence
of large extra dimensions; in these models, gravity could become strong at energies 
much lower than $M_{\rm Pl}$, and possibly accessible in future colliders~\cite{extra}.
However, these scenarios find encounter serious problems when they have to explain the 
cosmology of the early Universe. A third idea is to detect photons from very distant
sources and check if there is a delay in the arrival time of photons with different 
energies, as a consequence of Planck scale suppressed corrections to the standard 
dispersion relation~\cite{grbs}. It is not really clear if this is actually a test of quantum 
gravity or of the structure of the spacetime, and as of now all the data are consistent 
with the normal dispersion relation of special relativity.

This paper is the first study of a program whose aim is to investigate the possibility 
of observing quantum gravity related phenomena in the gravitational collapse of 
very massive stars. General relativistic equations for gravitational collapse can 
describe the final stages of the life of a star when its dense core implodes under 
its own gravity. In the standard picture, if the neutron degeneracy pressure threshold 
is passed, there is nothing capable of halting the collapse, and the final product is
a spacetime singularity, where the matter density diverges, predictability is lost, and
standard physics breaks down. Depending on the formation of trapped surfaces, the
singularity may either be hidden behind a horizon, and in this case the outcome of
the collapse would be a black hole, or be naked, and thus be visible to distant 
observers. The weak cosmic censorship conjecture asserts that singularities 
formed from gravitational collapse must be hidden within black holes~\cite{wccc}.
Although some examples are known in which naked singularities can form 
from regular initial data, their stability and genericity are not well understood at 
present; see e.g Ref.~\cite{ns} for some early results, and Ref.~\cite{review} 
for a recent review. We know that a black hole in which the central singularity 
is replaced by a finite distribution of exotic matter can in principle lose its 
horizon~\cite{zilong}, a result that suggests how the behavior of matter fields 
in the last stages of collapse is important for the horizon structure and hints toward 
a possible resolution of the central singularity. For some references on the 
possibility of observationally testing the existence of singularities, see 
e.g.~\cite{tests} and references therein. For all these reasons, it has now become 
of crucial importance to understand if we can observationally test the strong field 
regime where the classical relativistic framework might fail. Our strategy is to 
investigate how quantum effects can affect the formation of the singularity and 
of the trapped surfaces in order to see if the properties of the radiation emitted 
during the collapse in the ultra-dense region close to the classical singularity can 
reach distant observers and possibly carry information about the quantum gravity 
regime.

If we understand the singularity as a regime where the classical description breaks 
down and Planck scale effects arise, then the classical formation of a naked 
singularity suggests that these effects might propagate and influence the outside 
universe. The singularity would presumably be resolved within a theory of quantum 
gravity~\cite{Mode,Nico} and the existence of solutions where the classical singularity 
is naked would then indicate that spacetime regions of extremely high density occurring 
at the core of the collapse might be causally connected to the outside universe, 
thus bringing hints of new physics from catastrophic astrophysical events. Since 
we do not yet have any theory of quantum gravity, most attempts reduce to the 
study of some effective theory that describes quantum-gravity at a semiclassical 
level by the introduction of an appropriate modification of Einstein gravity in the 
strong field regime. Attempts to reformulate classical models within a quantum 
gravity approach have been successfully applied to cosmology~\cite{cosmology}, 
where the Big Bang singularity can be resolved and replaced by a bounce. In 
the context of gravitational collapse, loop quantum gravity (LQG) has been used 
to show how the singularity appearing at the end of the collapse can be 
removed~\cite{rbh}. It can be shown that quantum corrections to Einstein's 
equations can be put in a semiclassical framework where effective quantities 
take the place of the classical ones. Such an approach was used in Ref.~\cite{tavakoli},
where LQG corrections to the collapse of a scalar field were considered. Here, we 
will implement a similar strategy for the collapse of both a cloud of noninteracting particles 
(dust) and a perfect fluid with a linear equation of state that describes radiation. 
The treatment is completely classical and the quantum corrections appear 
in the form taken by the effective density and effective pressure of the system.

The Oppenheimer-Snyder (OS) marginally bound collapse of a dust sphere is 
the simplest case of gravitational collapse~\cite{os}. Although the model is 
extremely simple, it can give some insights into the behavior of classical gravity 
in the strong field regime. In this model, the singularity that forms at the end of 
the collapse is always hidden behind a horizon. It is known that the introduction 
of pressures in dust collapse can halt the process and cause a bounce. In the 
classical picture, in order to have a physically viable model, the pressure profile 
must satisfy certain assumptions, like the weak energy condition. On the other 
hand, we can study the case of collapse with pressures that lead to a bounce, 
neglecting the weak energy condition, if we are willing to reinterpret the ``exotic'' 
matter content on the right hand side of Einstein's equations as a semiclassical 
limit coming from an effective theory of quantum gravity inducing corrections 
in the small-scale/strong-gravity regime.

In this scenario, there is a new scale introduced in the evolution. It is governed 
by the value of the classical critical density $\rho_{\rm cr}$, which is a parameter 
{\it a posteriori} related to the Planck scale regime to be introduced from 
external considerations (such as LQG) and we can retrieve the classical solution 
in the limit of $\rho_{\rm cr}$ going to infinity. In our simplest quantum-gravity 
inspired gravitational collapse, the physical matter density reaches its maximum 
value $\rho_{\rm cr}$ at the critical time $t_{\rm cr} < t_{\rm s}$, where $t_{\rm s}$ is 
the time at which the singularity was reached in the classical case. At $t_{\rm cr}$,
the gravitational force is turned off (the effective density vanishes) and we are in a
regime of asymptotic freedom. At this time, we have a bounce and the collapsing
object starts expanding. As we consider the simplest case of homogeneous density
and pressure, at the bounce the gravitational force is switched off everywhere
and there is no apparent horizon. The latter forms again after the bounce and
eventually disappears forever. So, an event horizon never forms, in the sense 
that there are no regions causally disconnected from future null infinity. The collapse 
cannot form a black hole, but only a temporary apparent horizon that can mimic
a black hole for a time much shorter than the whole process of collapse and
expansion.

The paper is organized as follows. In Section~\ref{collapse}, we briefly summarize 
the classical framework for relativistic collapse:  in Subsection~\ref{dust}, we
review the basic equations of the OS homogeneous dust model, while 
Subsection~\ref{radiation} is for the Friedmann-Robertson-Walker (FRW) radiation 
collapse model. In Section~\ref{quantum}, we present our quantum-inspired 
gravitational collapse toy model and we show how the effective density coming 
from quantum corrections can resolve the formation of the singularity. Finally, 
Section~\ref{conclusion} is devoted to a brief summary and future perspectives.

\section{Gravitational collapse}\label{collapse}

The most general spherically symmetric metric describing a collapsing cloud of 
matter in comoving coordinates is given by
\begin{equation}\label{eq1}
    ds^2=-e^{2\nu}dt^2+\frac{R'^2}{G}dr^2+R^2d\Omega^2 \; ,
\end{equation}
where $d\Omega^2$ represents the line element on the unit two-sphere and 
$\nu$, $R$, and $G$ are functions of $t$ and $r$. The energy-momentum tensor 
is given by 
\begin{equation}
T^\mu_\nu={\rm diag}\{\rho(r,t), p_r(r,t), p_\theta(r,t), p_\theta(r,t) \} \; ,
\end{equation} 
and Einstein's equations relate the metric functions to the matter content:
\begin{eqnarray}\label{p}
  p_r &=&-\frac{\dot{F}}{R^2\dot{R}} \; , \\ \label{rho}
  \rho&=&\frac{F'}{R^2R'} \; , \\ \label{nu}
  \nu'&=&2\frac{p_\theta-p_r}{\rho+p_r}\frac{R'}{R}-\frac{p_r'}{\rho+p_r} \; ,\\ \label{Gdot}
  \dot{G}&=&2\frac{\nu'}{R'}\dot{R}G \; ,
\end{eqnarray}
where the $'$ denotes a derivative with respect to $r$, and the $\dot{}$ denotes a 
derivative with respect to $t$. The function $F(r,t)$, which is proportional to the 
amount of matter enclosed within the shell labeled by $r$ at the time $t$, is called 
Misner-Sharp mass, and is given by
\begin{equation}\label{misner}
F=R(1-G+e^{-2\nu}\dot{R}^2) \; .
\end{equation}

The whole system has a gauge degree of freedom that can be fixed by setting the 
scale at a certain time. In the case of collapse, the usual prescription is that the 
area radius $R(r,t)$ is set equal to the comoving radius $r$ at the initial time 
$t_{\rm i}=0$, $R(r,0)=r$. We can then introduce a scale function $a$
\begin{equation}
R(r,t)=ra(r,t) \; ,
\end{equation}
that will go from 1, at the initial time, to 0, at the time of the formation of the singularity. 
The condition to describe collapse is thus given by $\dot{a}<0$. The regularity of the 
energy density at the initial time, as seen from Eq.~\eqref{rho}, requires that 
$F(r,t)=r^3M(r,t)$, with $M(r,t) = \sum_{n = 0}^{\infty} M_n r^n$.

In order to have a physically realistic collapse, one typically requires some 
conditions for the matter model. Usually the assumptions are the following:
\begin{enumerate}
\item The regularity of the initial data for density and pressure.
\item The absence of cusps at the center for the energy density (which implies 
that $M'(0,t)=0$).
\item The energy density does not increase from the center outwards at any given time.
\item The weak energy condition ($\rho \ge 0$, $\rho+p_r \ge 0$, and 
$\rho+p_\theta \ge 0$). 
\end{enumerate}
We know that the energy conditions are averaged classical inequalities that do not 
take into consideration the microscopic properties of the matter and are likely to 
be violated in the semiclassical quantum regime~\cite{visser}.

\subsection{Classical dust model}\label{dust}

Let us now consider the simplest case of dust collapse, known as
Lemaitre-Tolman-Bondi (or LTB) model, where $p_r=p_\theta=0$~\cite{ltb}.
In the LTB model, from Eq.~\eqref{p} one immediately gets that $M=M(r)$ and 
the cloud can be matched to a Schwarzschild exterior with total mass 
$2M_{\rm T}=F(r_{\rm b})$ at the boundary $r_{\rm b}$~\cite{matching}. From 
Eq.~\eqref{nu}, one can choose the time gauge in such a way that $\nu=0$. 
Then Eq.~\eqref{Gdot} implies $G=1+f(r)$, which in the marginally bound case 
(representing particles that fall from infinity with zero initial velocity) simply 
becomes $G=1$. The metric is then given by
\begin{equation}
ds^2=-dt^2+R'^2dr^2+R^2d\Omega^2 \; .
\end{equation}
The Misner-Sharp mass, Eq.~\eqref{misner}, takes the form of an equation of motion
\begin{equation}\label{motion}
\dot{a}=-\sqrt{\frac{M}{a}} \; ,
\end{equation}
with the minus sign chosen in order to describe a collapse. The integration of 
Eq.~\eqref{motion} is straightforward and gives
\begin{equation}\label{a1}
a(r,t)=\left(1-\frac{3}{2}\sqrt{M}t\right)^{2/3} \; .
\end{equation}
Then the remaining Einstein's equation, Eq.~\eqref{rho}, is written as
\begin{eqnarray}\label{rho2}
  \rho=\frac{3M+rM'}{a^2(a+ra')} \; ,
\end{eqnarray}
which completely solves the system.

The model has a strong curvature singularity for $a\rightarrow 0$, as can be seen
from the divergence of the Kretschmann scalar
\begin{eqnarray}
R_{\mu\nu\rho\sigma} R^{\mu\nu\rho\sigma}
=12 \frac{\ddot{a}^2 a^2 + \dot{a}^4}{a^4} \; .
\end{eqnarray}
The singularity is achieved along the curve $t_{\rm s}(r)= 2/3\sqrt{M}$ and 
the central line $r=0$ is regular for $a\neq 0$. The central singularity $t_{\rm s}(0)$ 
can be visible to faraway observers depending on the matter profile $M(r)$~\cite{dust}. 
In the present work, we restrict our attention to the simplest case, the OS dust 
collapse, where the choice of $M(r)=M_0$ causes the density to be homogeneous 
and the final outcome is a black hole. In the OS model, the metric takes the form
\begin{equation}\label{FRW}
ds^2=-dt^2+a^2(dr^2+r^2d\Omega^2) \; ,
\end{equation}
which is the time reversal of the FRW cosmological scenario. We then get
\begin{eqnarray}\label{kkk}
\rho(t)&=&\frac{3M_0}{a^3} \; , \\ \label{kkk2}
M_0&=&a\dot{a}^2\; , \\ \label{kkk3}
a(t)&=&\left(1-\frac{3}{2}\sqrt{M_0}t\right)^{2/3} \; , 
\end{eqnarray}
and all the shells become singular at the same comoving time 
$t_{\rm s} = 2/3\sqrt{M_0}$. Here, the apparent horizon forms at the boundary 
at a time antecedent to the formation of the singularity, thus leaving the region 
where Planck scale effects arise hidden from faraway observers.

\subsection{Classical radiation model}\label{radiation}

We turn now to the classical FRW solution describing the collapse of a 
homogeneous perfect fluid, where $p_r=p_\theta=p(t)$, and consider the case 
of radiation where the equation of state relating the pressure to the density is 
\begin{equation}\label{eos}
\rho=3p \; .
\end{equation}
The isotropy and homogeneity of the pressure implies that Eq.~\eqref{Gdot} must 
give $G=1+f(r)$, in analogy with the dust case. Again, we will consider here the 
marginally bound solution with $f=0$. Unlike in the dust scenario, the presence of 
the homogeneous pressure indicates that the mass profile is not constant throughout 
the collapse, i.e. $M$ must depend on $t$, and therefore the matching with the exterior 
must be done with the Vaidya solution~\cite{santos}. The equation of state together 
with Einstein's equations~\eqref{p} and \eqref{rho} implies that the mass profile 
must satisfy the following differential equation
\begin{equation}
\frac{dM}{da}=-\frac{M}{a} \; ,
\end{equation}
which gives $M=\frac{M_0}{a}$. Then the energy density becomes 
\begin{equation}\label{kkkk}
\rho=\frac{3M_0}{a^4} \; ,
\end{equation}
and
\begin{equation}\label{kkkk2}
M_0=a^2\dot{a}^2 \; .
\end{equation}
Finally, the integration of the equation of motion~\eqref{motion} gives
\begin{equation}
a(t)=(1-2\sqrt{M_0}t)^{1/2} \; .
\end{equation}
Once again, the metric is given by Eq.~\eqref{FRW}, and the singularity 
occurs at the same time $t_s=1/2\sqrt{M_0}$ for each shell. Like in 
the dust model, the final outcome is a black hole.

\section{Quantum-inspired collapse}\label{quantum}

The above system of Einstein's equations for dust or radiation collapse is closed 
once a free function is specified -- typically the mass profile $M$ or the density 
profile $\rho$. In the case of the OS model, we have chosen the mass profile 
$M=M_0$, while for the radiation model we have specified the equation of 
state~\eqref{eos}. Therefore an effective model of quantum gravitational collapse 
can be given by a well motivated choice of the free function that replaces the 
classical choice, while introducing a scale factor in the form of a critical density 
$\rho_{\rm cr}$ that can be related to the Planck scale. We can then interpret 
the model as a modification to the standard dust or radiation collapse scenario 
induced by quantum corrections in the strong field limit. To this aim, we can 
rewrite the right-hand side of Einstein's equations as dust + corrections or 
radiation + corrections, and the newly introduced parameter $\rho_{\rm cr}$ 
indicates the scale at which the corrections become relevant. Typically, we will 
have the correction becoming important at high densities, so one can write
\begin{equation}\label{corr}
\rho_{\rm corr}= \alpha_1 \rho^2
+ \alpha_2 \rho^3 + o(\rho^3) \; ,
\end{equation}
where the parameters $\alpha_i=\alpha_i(\rho_{\rm cr})$ that determine the 
scale of the quantum corrections will go as an inverse power of $\rho_{\rm cr}$ 
and in general will be determined from the quantum theory. In this way, we can 
write $T_{\mu\nu}=T_{\mu\nu}^{\rm class} +T_{\mu\nu}^{\rm corr}$, where, for 
dust, we obviously have $p_{\rm class}=0$. We can then move the correction 
$T_{\mu\nu}^{\rm corr}$ to the left-hand side of Einstein's equations, thus reinterpreting 
the model as a dust or radiation solution in some effective theory of gravity that 
accounts for corrections in the strong field limit. This procedure is completely 
equivalent to replacing Newton's constant for classical gravity $G_{\rm N}$ 
with a variable coupling function $G(\rho)$ derived from the effective quantum 
theory.

We expect that quantum effects become relevant towards the formation of the 
singularity, as they are supposed to ``smear'' the singularity, thus avoiding the 
breakdown of predictability that occurs in the classical case. In general, the 
system of Einstein's equations for perfect fluid collapse, when the equation 
of state is not specified, leaves the freedom to choose one free function. It is not 
difficult to show that a suitable choice of the mass profile can drastically change 
the structure underlying the formation of the horizon and singularity (see, for example, 
Ref.~\cite{pjm}). In order to study how the inclusion of quantum effects in the 
form of an effective theory affects the formation of the singularity at the end of 
the collapse, we assume that at first order the corrections due to quantum gravity 
take the form given by Eq.~\eqref{corr} and thus we guess an effective energy 
density profile as a function of the dust and radiation energy densities (as written 
in Eqs.~\eqref{kkk} and \eqref{kkkk}, respectively) as $\rho_{\rm eff}=\rho
+\rho_{\rm corr}$ and take this as the free function for the system.

Following Ref.~\cite{tavakoli}, we consider here an effective theory of gravity 
where the corrections to the energy density~\eqref{corr} take the form
\begin{equation}\label{rho-eff0}
\rho_{\rm eff}=\rho\left(1-\frac{\rho}{\rho_{\rm cr}}\right)^\gamma, 
\; \gamma\geq 1 \; .
\end{equation}
and the effective density specified by this equation will play the role of the free 
function in the effective model for collapse. In the following, we will consider 
the cases $\gamma=1$ and 2. The case $\gamma=1$ corresponds to the 
choice of $\alpha_1=- 1/\rho_{\rm cr}$ and $\alpha_i=0$ for $i>1$. The 
case $\gamma=2$ corresponds to the choices $\alpha_1=- 2/\rho_{\rm cr}$, 
$\alpha_2=1/\rho_{\rm cr}^2$, and $\alpha_i=0$ for $i>2$. In the weak 
field limit, for large $\rho_{\rm cr}$, quantum corrections become negligible, 
and in the limit of infinite $\rho_{\rm cr}$ we recover the classical dust and 
radiation cases. For low densities (i.e. close to the initial time), the effective 
energy density approaches that of the classical models.

\subsection{Quantum-inspired dust model}\label{quantumdust}

Let us note that the scale function $a$ that appears in Eq.~\eqref{kkk} must 
now be determined from the integration of the new equation of motion coming 
from Eq.~\eqref{corr} that replaces Eq.~\eqref{kkk2}. This now becomes
\begin{equation}
\dot{a}^2=\frac{M_0}{a}+\alpha_1\frac{3M_0^2}{a^4}
+\alpha_2\frac{9M_0^3}{a^7}+...
\end{equation}

Here, we will consider an effective density of the form given in Eq.~\eqref{rho-eff0},
as suggested by first order corrections coming from LQG, and integrate for 
the cases $\gamma=1$ and $2$. We then proceed to solve Einstein's equations 
for the effective theory where the density $\rho$, the pressure $p$, 
and the Misner-Sharp mass $M$ are replaced by the 
corresponding effective quantities (namely $\rho_{\rm eff}$, $p_{\rm eff}$, and 
$M_{\rm eff}$). Eq.~\eqref{kkk} then becomes
\begin{equation}\label{rho-eff}
\rho_{\rm eff}=\frac{3M_{\rm eff}}{a^3} \; ,
\end{equation}
where the new effective mass $M_{\rm eff}(t)$ is again given by $M_{\rm eff}(t)=
a\dot{a}^2$ and is not constant (the fact that $M_{\rm eff}$ is a function of $t$ 
induces the presence of the effective pressure $p_{\rm eff}$). We then obtain 
the differential equation for the scale function $a(t)$ from Eqs.~\eqref{rho-eff0} 
and \eqref{rho-eff}: 
\begin{equation}\label{diff}
\dot{a}^2=\frac{M_0}{a^{3\gamma+1}}\left(a^3-a_{\rm cr}^3\right)^\gamma, 
\;\; \text{with} \;\;
a_{\rm cr}^3=\frac{3M_0}{\rho_{\rm cr}} \;,
\end{equation}
where we have introduced the critical scale $a_{\rm cr}$. Let us note that, for 
$a_{\rm cr}\rightarrow 0$, we recover the dust solution with $\rho_{\rm eff}=\rho$ 
and $a(t)$ given by Eq.~\eqref{kkk3}. Once solved with the initial condition 
$a(1)=1$, Eq.~\eqref{diff} gives
\begin{widetext}
\begin{eqnarray}
t(a)&=& \frac{2}{3\sqrt{M_0}}\left(\sqrt{1-a_{\rm cr}^3}-
\sqrt{a^3-a_{\rm cr}^3}\right)
\; \; \text{for $\gamma=1$} \; , \\
t(a)&=& \frac{2}{3\sqrt{M_0}}(1-a^{3/2})-\frac{a_{\rm cr}^{3/2}}{3\sqrt{M_0}}
\ln{\frac{(a^{3/2}-a_{\rm cr}^{3/2})(1+a_{\rm cr}^{3/2})}{(a^{3/2}+a_{\rm cr}^{3/2})(1-a_{\rm cr}^{3/2})}}
\; \; \text{for $\gamma=2$} \; .
\end{eqnarray}
\end{widetext}
The metric for the effective model is still described by the usual form given by 
Eq.~\eqref{FRW}.

\begin{figure*}
\begin{center}
\hspace{-0.6cm}
\includegraphics[type=pdf,ext=.pdf,read=.pdf,width=7.75cm]{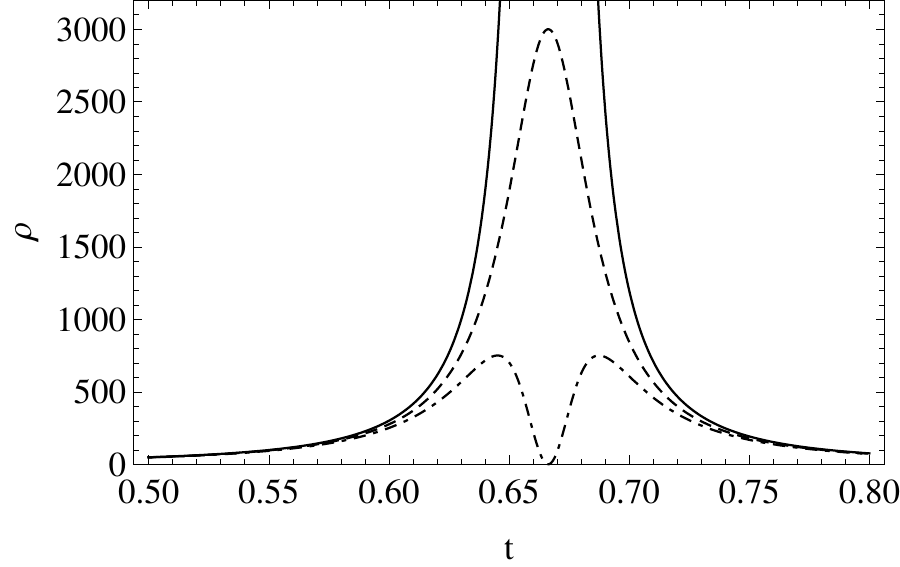} 
\hspace{1.2cm}
\includegraphics[type=pdf,ext=.pdf,read=.pdf,width=7.7cm]{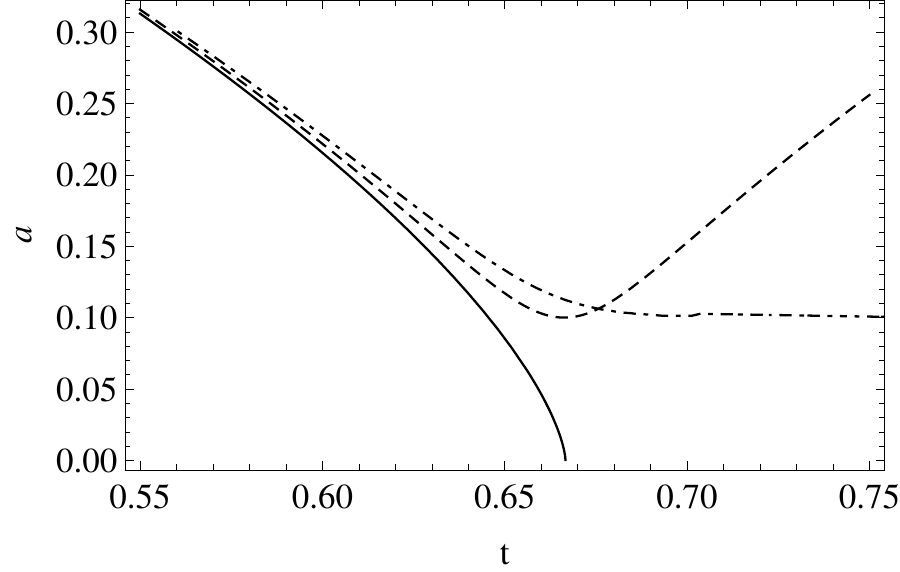}
\end{center}
\caption{Dust collapse models. Left panel: the density $\rho$ in the classical 
model (solid line), the density $\rho$ in the quantum-gravity inspired collapse
model with $\gamma = 1$ (dashed line), and the effective density $\rho_{\rm eff}$
in the quantum-gravity inspired collapse model with $\gamma = 1$ (dashed-dotted 
line). Right panel: plot of $a(t)$ in the classical case (solid line) and in the 
semiclassical model with $\gamma=1$ (dashed line) and $\gamma=2$ (dotted-dashed 
line). Near the initial time, the semiclassical model has a behavior close to 
the dust. $a$ either reaches a minimum at $t = t_{\rm cr}$ and then grows 
for $t>t_{\rm cr}$ ($\gamma = 1$), or approaches asymptotically a minimum value
($\gamma = 2$). Here, $M_0 = 1$ and $\rho_{\rm cr} = 3000$. See the text for details.}
\label{fig1}
\vspace{0.8cm}
\begin{center}
\hspace{-0.4cm}
\includegraphics[type=pdf,ext=.pdf,read=.pdf,width=7.5cm]{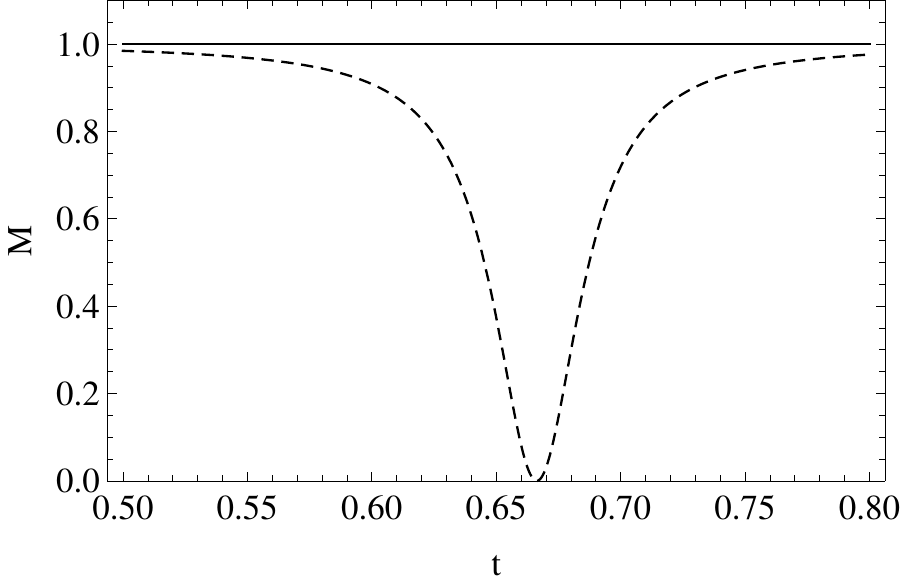} 
\hspace{0.8cm}
\includegraphics[type=pdf,ext=.pdf,read=.pdf,width=8cm]{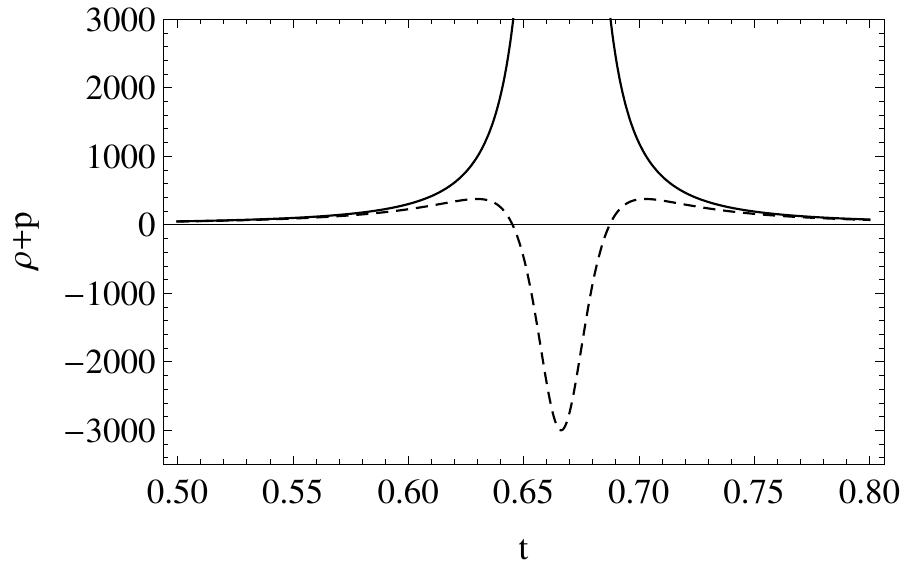}
\end{center}
\caption{Dust collapse models. Left panel: the mass profile $M=M_0$ (solid line) 
and $M_{\rm eff}$ for $\gamma=1$ (dashed line). At $t = t_{\rm cr}$, the effective 
mass vanishes and thus the spacetime is flat; we are in a regime of asymptotic 
freedom. Right panel: in order to investigate the breakdown of the weak energy 
condition, we plot $\rho$ for the classical dust case (solid line), and $\rho_{\rm eff}
+p_{\rm eff}$ for $\gamma=1$ (dashed line). Here, $M_0 = 1$ and $\rho_{\rm cr} = 3000$. 
See the text for details.}
\label{fig2}
\end{figure*}

The effective energy density for the toy model studied here is homogeneous. 
This causes the effective dynamics of the system to be equivalent to that of the 
collapse of a homogeneous perfect fluid where the pressure, coming from 
Einstein's equation \eqref{p}, is given by
\begin{equation}\label{peff}
p_{\rm eff}(t)=-\frac{\dot{M}_{\rm eff}}{a^2\dot{a}} \; .
\end{equation}
The effective pressure that describes the quantum corrections in the semiclassical 
theory is then homogeneous as well. From 
\begin{equation}\label{Meff}
M_{\rm eff}=M_0\left(1-\frac{\rho}{\rho_{\rm cr}}\right)^\gamma \; ,
\end{equation} 
we see that the effective pressure becomes
\begin{equation}
p_{\rm eff}=-\gamma\frac{\rho^2}{\rho_{\rm cr}}
\left(1-\frac{\rho}{\rho_{\rm cr}}\right)^{\gamma-1} \; .
\end{equation}
The effective pressure is always negative and, in the case $\gamma=1$, approaches 
the limiting case of $p=-\rho$ as the density approaches $\rho_{\rm cr}$. On the other 
hand, in the case $\gamma>1$, the pressure goes to zero as $\rho \rightarrow 
\rho_{\rm cr}$ (as expected, it becomes zero in the limit of $\rho_{\rm cr}$ going to 
infinity). The fact that the effective pressure becomes negative in the strong field 
regime is what causes the bounce of the new density. At the initial time, the new 
density is equal to the density in the classical case, while the effective pressure is 
very small, although not zero (being zero in the limit of $\rho_{\rm cr}$ going to 
infinity). From these considerations, it is easy to verify that the weak energy 
condition is satisfied at the beginning of collapse, and it is violated as the quantum 
gravity regime is approached.

It is worth noting that the scale function $a(t)$ behaves differently for the models 
with $\gamma=1$ and $\gamma=2$. In the first case, $a$ reaches the minimum 
value $a_{\rm cr}$ in a finite time $t_{\rm cr}$, and then grows indefinitely, thus 
originating a gravitational bounce. On the other hand, in the case $\gamma=2$, 
the minimum at $a_{\rm cr}$ is reached only as $t \rightarrow \infty$ and therefore 
there is no bounce but an indefinite collapse that slows down until it stops asymptotically 
(see the right panel in Fig.~\ref{fig1}). The different behavior of $a$ for different 
values of $\gamma$ is also reflected in a different behavior for the apparent horizon. 
In the case $\gamma=1$, the apparent horizon curve $r_{\rm ah}(t)$ diverges for 
$t$ approaching $t_{\rm cr}$, while, in the case $\gamma=2$, it diverges for 
$t \rightarrow \infty$.

We will now concentrate on the case $\gamma=1$. The scale function has the form
\begin{equation}\label{v}
a(t)= \left[a_{\rm cr}^3+\left(\sqrt{1-a_{\rm cr}^3}-
\frac{3\sqrt{M_0}}{2}t\right)^2\right]^{1/3} \; ,
\end{equation}
and it reaches a minimum at the time 
\begin{equation}
t_{\rm cr}=\frac{2\sqrt{1-a_{\rm cr}^3}}{3\sqrt{M_0}}<t_{\rm s},
\end{equation}
 at which $\rho_{\rm eff}$ vanishes, and then increases for
$t>t_{\rm cr}$ (see the right panel in Fig.~\ref{fig1}). The fact that the time of the 
bounce $t_{\rm cr}$ occurs before the classical time of the singularity is due to 
the choice of the integration constant in Eq.~\eqref{v}, which for collapse is 
chosen in order to have $a(0)=1$. In standard cosmological models, the constant 
is chosen in such a way that $t_{\rm s}=t _{\rm cr}$ and leads to an initial condition 
$a^3(0)=1+a_{\rm cr}^3>1$ for the effective model. Of course, the metric is invariant 
under time translations, so the crucial point here is that once an initial time $t_{\rm i}$ 
is fixed in such a way that the metric coefficients take the same numerical values 
in the classical and quantum case then the interval $\Delta t_{\rm s}=|t_{\rm s}-t_{\rm i}|$ 
is greater than the interval $\Delta t_{\rm cr}=|t_{\rm cr}-t_{\rm i}|$. The effective 
density reaches a maximum at the time 
$t_{\rm max}=t_{\rm s}(\sqrt{1-a_{\rm cr}^3}-\sqrt{a_{\rm cr}^{3}})$ 
and then decreases, becoming zero at the critical time $t_{\rm cr}$.

Contrary to the classical dust case, where $\rho$ diverges at the time 
$t_{\rm s}=2/3\sqrt{M_0}$, in our case we can see that the new density 
$\rho=3M_0/a^3$ tends to the maximum value $\rho_{\rm cr}$ as 
$t$ goes to $t_{\rm cr}$ and then decreases (see the left panel in Fig.~\ref{fig1}). 
Also, the velocity of the collapsing shells $\dot{a}$ tends to zero as $t$ goes to 
$t_{\rm cr}$, unlike the classical case where the velocity is finite at $t_{\rm cr}$ 
and diverges when we approach the singularity.

In this model, the strong curvature singularity is removed: the Kretschmann 
scalar is still given by Eq.~\eqref{kkk}, but with the new scale function $a$ it never 
diverges. Furthermore, the fact that we are dealing with a homogeneous perfect 
fluid for which the area radius is $R(r,t)=ra(t)$, together with the positivity of 
the scale function $a$, ensures that there are no shell crossing singularities in 
the spacetime, which is therefore everywhere regular until $t=t_{\rm cr}$ and can 
be prolonged for $t>t_{\rm cr}$. The effective mass of the collapsing perfect fluid 
cloud is given by Eq.~\eqref{Meff}. We can see that $M_{\rm eff}$ decreases, 
becoming zero in the limit of $t$ going to $t_{\rm cr}$ (see the left panel in 
Fig.~\ref{fig2}). Therefore the matching with the exterior spacetime must be done 
with the Vaidya solution describing outgoing radiation. Further, from the fact that 
$\rho_{\rm eff}(t_{\rm cr})=M_{\rm eff}(t_{\rm cr})=0$, at $t=t_{\rm cr}$ 
the spacetime is flat. This happens because in our model gravity becomes
weaker and weaker as $\rho$ approaches $\rho_{\rm cr}$ and is turned off
when $\rho = \rho_{\rm cr}$. The bounce occurs at $t=t_{\rm cr}$ and then the
collapse changes into an expansion. At lower densities, we recover Einstein's
gravity, but now the model describes an expanding cloud with $\dot{a}>0$.

The collapsing matter is usually required to satisfy the weak energy condition, 
which in the case of a perfect fluid reads $\rho+p \ge 0$. For the effective theory 
described here, it is easy to check that $\rho_{\rm eff}+p_{\rm eff} \ge 0$ is 
satisfied in the weak field regime, close to the initial time, while it is violated due 
to the negative pressures as we approach the critical density $\rho_{\rm cr}$
(see the right panel in Fig.~\ref{fig2}).

\begin{figure}
\begin{center}
\vspace{0.2cm}
\hspace{-0.6cm}
\includegraphics[type=pdf,ext=.pdf,read=.pdf,width=7.5cm]{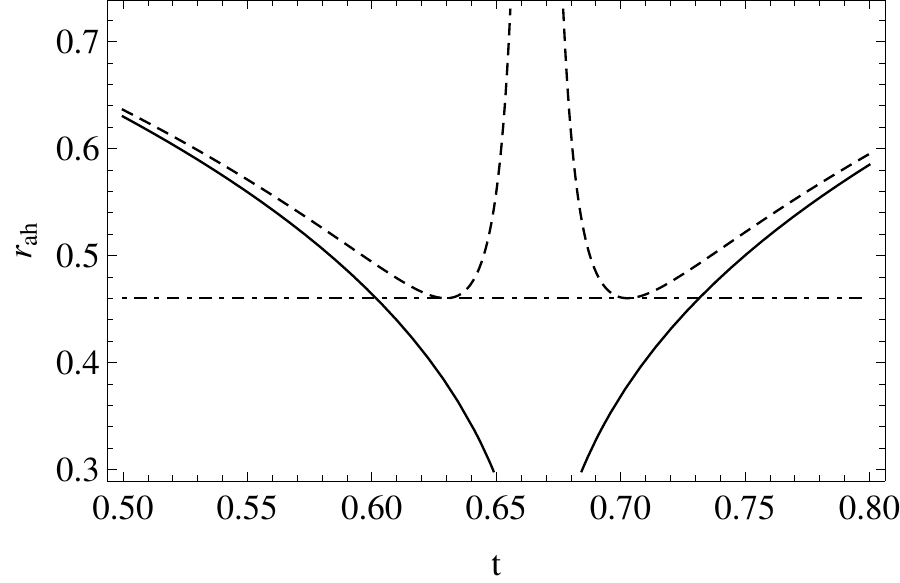}
\end{center}
\caption{The apparent horizon curve $r_{\rm ah}(t)$ for the classical dust 
model (solid line) and the semiclassical model for $\gamma=1$ (dashed line). 
If the boundary of the cloud is taken smaller than $r_{\rm min}$ (dashed-dotted 
line) there are never trapped surfaces forming in the quantum inspired model.
At the time of the bounce, when $\rho = \rho_{\rm cr}$ and gravity is turned off,
the spacetime is flat, and there is no horizon. This is a regime of asymptotic 
freedom.}
\label{fig3}
\end{figure}

In general relativity, under common assumptions like matter energy conditions
and cosmic censorship, the appearance of an apparent horizon is related to the
existence of an event horizon. This is not the case here, because of the unconventional
properties of our effective matter. Despite that, it is convenient to study the 
apparent horizon of the spacetime and to compare the results with the scenario
of the standard picture.
The condition for the formation of trapped surfaces is given by the requirement 
that the surface $R(r,t) = {\rm constant}$ is null; that is, $g^{\mu\nu} (\partial_\mu R)
(\partial_\nu R) = 0$. For the metric in Eq.~\eqref{eq1}, this means 
\begin{equation}
G - e^{-2\nu} \dot{R}^2 = 0 \; ,
\end{equation}
and, from the definition of Misner-Sharp mass~\eqref{misner}, we can write it as
\begin{equation}
1-\frac{F}{R}=0 \; .
\end{equation}
In the dust case, the condition for the absence of trapped surfaces at the initial 
time reduces to $r^2M_0<1$, while in the model presented here we must require 
$r^2M_0\sqrt{1-a_{\rm cr}^3}<1$. In the classical OS case, the apparent horizon 
curve is given by 
\begin{equation}
t_{\rm ah}(r)=t_{\rm s}-\frac{2}{3}r^3M_0 \; .
\end{equation}
In our model, the 
central singularity is avoided. What happens to the formation of trapped surfaces? 
The equation for the apparent horizon becomes
\begin{equation}
r_{\rm ah}(t)=\frac{a^2}{\sqrt{M_0(a^3-a_{\rm cr}^3)}} \; .
\end{equation}
It is not difficult to check that $r_{\rm ah}$ has a minimum for 
\begin{equation}
t=t_{\rm min}=t_{\rm s}(\sqrt{1-a_{\rm cr}^3}-\sqrt{3a^3_{\rm cr}}).
\end{equation}
Therefore there exists a minimum radius 
\begin{equation}
r_{\rm min}=r_{\rm ah}(t_{\rm min})=2^{4/3}\sqrt{\frac{a_{\rm cr}}{3M_0}} \; ,
\end{equation} 
for which, if the boundary is taken as $r_{\rm b}<r_{\rm min}$, no trapped surfaces 
form during the whole process of collapse and bounce (see Fig.~\ref{fig3}).

The existence of a minimum radius implies that in the quantum modified scenario 
there exists a minimal mass $M_{\rm min}$ below which an apparent horizon
never forms. In fact, considering the boundary condition for dust collapse 
$2M_T=r_{\rm b}^3M_0$, where $M_T$ is the total mass in the exterior spacetime, 
if we take the boundary $r_{\rm b}=r_{\rm min}$, we can evaluate $M_{\rm min}$ as
\begin{equation}
M_{\rm min}=8\sqrt{\frac{a_{\rm cr}^3}{27M_0}} \; .
\end{equation}
From the above equation, we see that, if the critical density is taken of the order of 
the Planck density, then $M_{\rm min}$ must be very small.

In the more general case for larger collapsing objects, an apparent horizon forms
at a time $t$ a little bit later than the classical case. However, when the density $\rho$
approaches the critical density $\rho_{\rm cr}$, gravity becomes more and more 
weak (it is turned off completely when $\rho = \rho_{\rm cr}$, which occurs at the
time $t = t_{\rm cr}$) and the spacetime reduces to the flat Minkowski case. The
apparent horizon thus disappears ($r_{\rm ah}$ diverges) and the bounce is 
``immediately'' visible to distant observers. 
Let us notice, however, that our model assumes a homogeneous 
density and therefore at the critical time $t_{\rm cr}$ gravity is turned off everywhere.
In a more realistic scenario, where the density is higher at the center,
the bounce may still be hidden behind a horizon produced by the matter at larger
radii and lower densities. After the bounce, the collapse is reversed into an expansion 
and the matter density starts decreasing. As the asymptotic freedom regime is
left, gravity becomes strong again and a new apparent horizon forms. However,
we are now in an expanding phase, and when the matter density becomes too
low, the apparent horizon disappears forever. 
Since we are considering the marginally bound case, where collapse has
zero velocity at spatial infinity, in the expanding phase the cloud will return 
to its initial configuration, but with positive velocity, and continue
to expand until all the matter is radiated to infinity.

\vspace{0.5cm}

\subsection{Quantum-inspired radiation model}\label{quantumradiation}

\begin{figure*}
\begin{center}
\hspace{-0.3cm}
\includegraphics[type=pdf,ext=.pdf,read=.pdf,width=8cm]{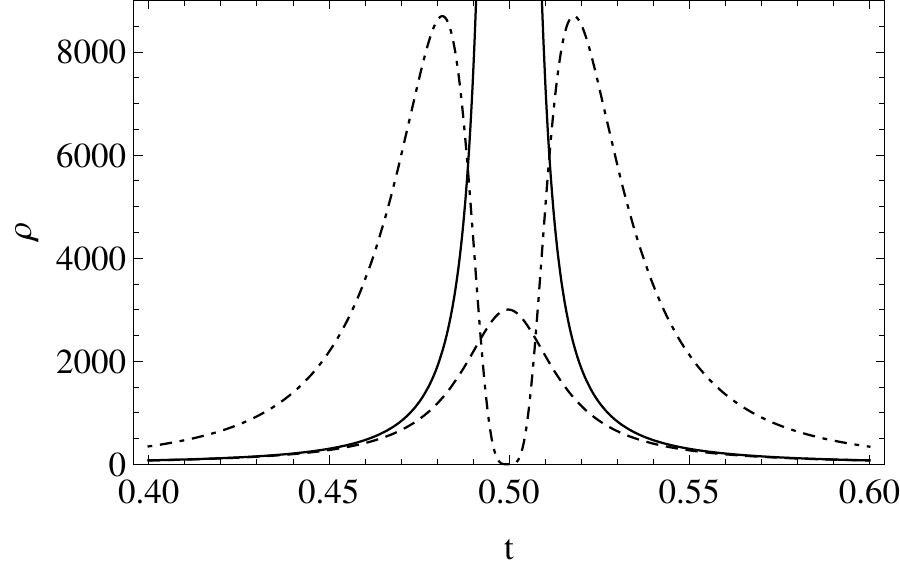} 
\hspace{1cm}
\includegraphics[type=pdf,ext=.pdf,read=.pdf,width=7.65cm]{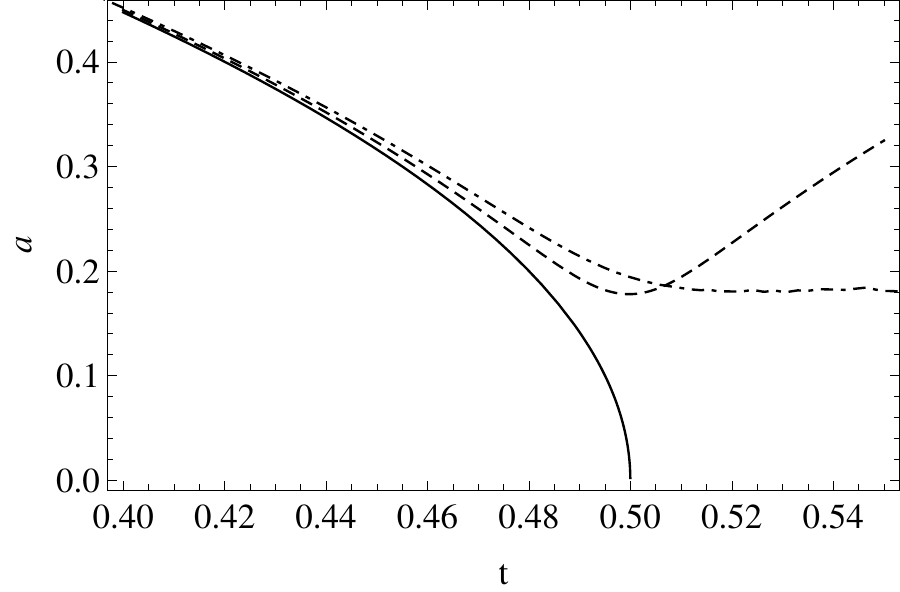}
\end{center}
\caption{Radiation collapse models. Left panel: the density $\rho$ in the classical
model (solid line), the density $\rho$ in the quantum-inspired collapse model with
$\gamma = 1$ (dashed line), and the effective density $\rho_{\rm eff}$ in the 
quantum-inspired collapse model with $\gamma = 1$ (dashed-dotted line). 
Right panel: plot of $a(t)$ in the classical case (solid line) and in the semiclassical 
models with $\gamma=1$ (dashed line) and $\gamma=2$ (dotted-dashed line).
Near the initial time, the semiclassical model has a behavior close to the classical
radiation. $a$ either reaches a minimum at $t = t_{\rm cr}$ and then grows for 
$t>t_{\rm cr}$ ($\gamma = 1$), or approaches asymptotically a minimum value 
($\gamma = 2$). Here, $M_0 = 1$ and $\rho_{\rm cr} = 3000$. See the text for details.}
\label{fig4}
\vspace{0.8cm}
\begin{center}
\hspace{-0.3cm}
\includegraphics[type=pdf,ext=.pdf,read=.pdf,width=7.6cm]{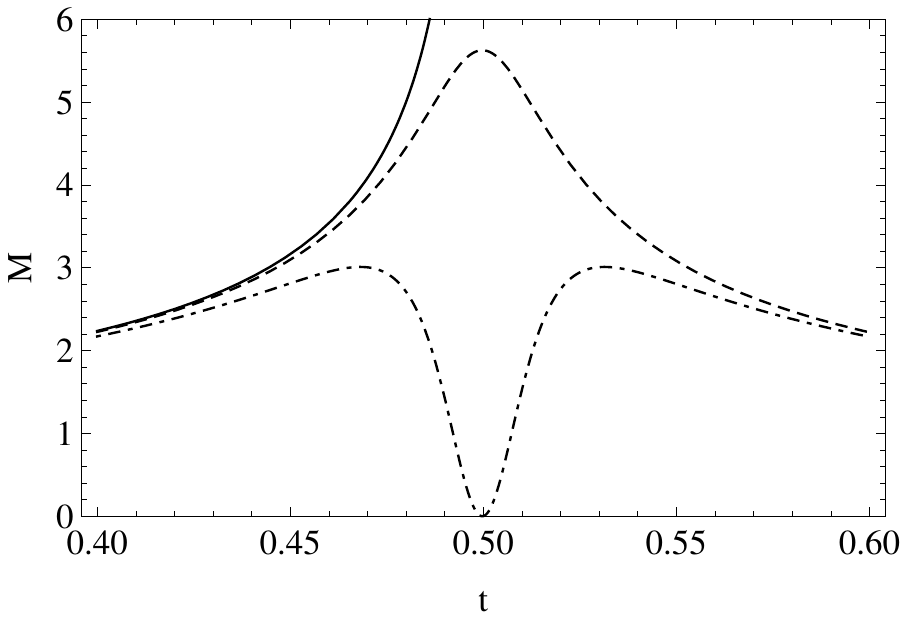} 
\hspace{1cm}
\includegraphics[type=pdf,ext=.pdf,read=.pdf,width=8.3cm]{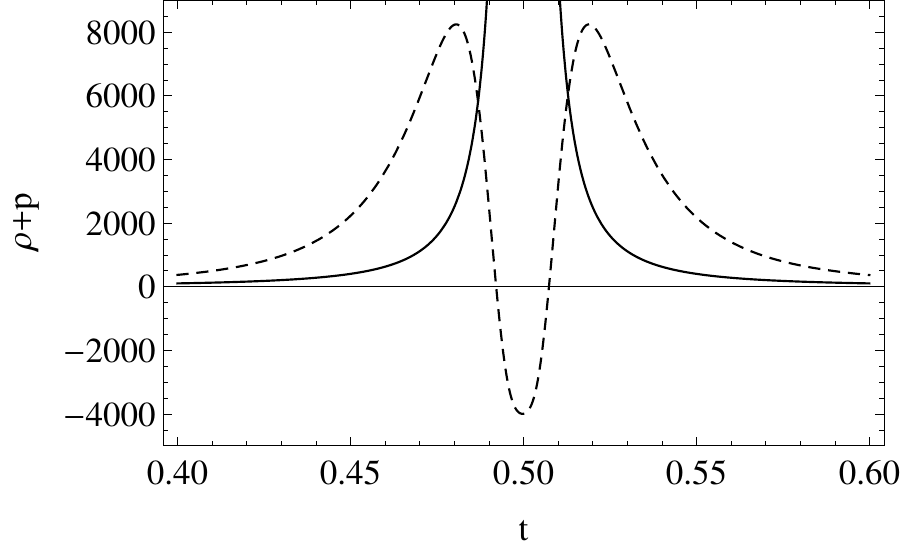}
\end{center}
\caption{Radiation collapse models. Left panel: the mass profile $M(t)=M_0/a$ 
for classical radiation (solid line) and for the semiclassical model (dashed line), 
together with the effective mass profile $M_{\rm eff}$ (dotted-dashed line). At 
$t = t_{\rm cr}$, in the semiclassical model $M(t)$ reaches a maximum, while the effective 
mass vanishes: the spacetime is flat and we are in a regime of asymptotic freedom. 
Right panel: $\rho+p$ for the classical radiation case (solid line) and 
$\rho_{\rm eff}+p_{\rm eff}$ for $\gamma=1$ (dashed line). In the latter case, the 
weak energy condition does not hold in the strong field limit. Here, $M_0 = 1$ and 
$\rho_{\rm cr} = 3000$. See the text for details.}
\label{fig5}
\end{figure*}

Following what we did for the dust case, we turn now to the radiation 
collapse model, where the new scale function $a$ that appears in Eq.~\eqref{kkkk} 
has to be determined from the integration of the new equation of motion
\begin{equation}
\dot{a}^2=\frac{M_0}{a^2}+\alpha_1\frac{3M_0^2}{a^6}
+\alpha_2\frac{9M_0^3}{a^{10}}+... \; .
\end{equation}
The effective density is again taken to be of the form given in Eq.~\eqref{rho-eff} and 
the equation of motion becomes
\begin{equation}
\dot{a}^2=\frac{M_0}{a^{4\gamma+2}}(a^4-a_{\rm cr}^4)^\gamma,
\;\; \text{with} \;\; a_{\rm cr}^4=\frac{3M_0}{\rho_{\rm cr}} \; .
\end{equation}
Solving the above equation in the two cases $\gamma=1$ and 2, with the initial condition 
$a(0)=1$, we obtain
\begin{widetext}
\begin{eqnarray}
t(a)&=&\frac{\sqrt{1-a_{\rm cr}^4}-\sqrt{a^4-a_{\rm cr}^4}}{2\sqrt{M_0}} , 
\; \; \text{for $\gamma=1$} \; , \\
t(a)&=& \frac{1-a^2}{2\sqrt{M_0}}-\frac{a_{\rm cr}^2}{2\sqrt{M_0}}\left(\tanh^{-1}
\left(\frac{1}{a_{\rm cr}^2}\right)-\tanh^{-1}\left(\frac{a^2}{a_{\rm cr}^2}\right)\right),
\; \; \text{for $\gamma=2$} \; .
\end{eqnarray}
\end{widetext}
Once again, the two cases are substantially different since the scale function $a(t)$ 
reaches its minimum value $a_{\rm cr}$ in a finite time $t_{\rm cr}$ for $\gamma=1$, 
while it needs an infinite time when $\gamma=2$ (see the right panel in Fig.~\ref{fig4}).

The effective pressure is still given by Eq.~\eqref{peff}, where now we have the 
effective mass given by
\begin{equation}\label{Meff2}
M_{\rm eff}=\frac{M_0}{a}\left(1-\frac{\rho}{\rho_{\rm cr}}\right)^\gamma \; .
\end{equation} 
The effective mass goes to zero as $t$ approaches $t_{\rm cr}$, while the new 
mass for the radiation fluid, given by $M_0/a$, reaches a maximum value 
(see the left panel in Fig.~\ref{fig5}). Evaluating the effective pressure, we find
\begin{equation}
p_{\rm eff}=\frac{\rho}{3}\left(1-5\gamma\frac{\rho}{\rho_{\rm cr}}\right)
\left(1-\frac{\rho}{\rho_{\rm cr}}\right)^{\gamma-1} \; .
\end{equation}
In the strong field region, the effective pressure becomes negative. In the case 
$\gamma=1$, we see that $p_{\rm eff}$ becomes negative when the density 
reaches the value $\rho_{\rm cr}/5$, and it tends to $- 4\rho/3$ in the 
limit of $\rho \rightarrow \rho_{\rm cr}$. In the case $\gamma=2$, it becomes 
negative when the density is $\rho_{\rm cr}/10$, and then goes back to 
zero when $\rho$ approaches $\rho_{\rm cr}$. The weak energy condition for 
the effective dynamics is given by $\rho_{\rm eff}+p_{\rm eff}$ and it is violated 
in the strong field regime (see the right panel in Fig.~\ref{fig5}).

We will now focus on the case $\gamma=1$, for which the scale function 
becomes
\begin{equation}
a(t)=[a_{\rm cr}^4+(\sqrt{1-a_{\rm cr}^4}-2\sqrt{M_0}t)^2]^{1/4} \; ,
\end{equation}
$a$ reaches a minimum at $t_{\rm cr}<t_{\rm s}$, where $\dot{a}$ vanishes. At the 
critical time, the effective density goes to zero, while the new density reaches its 
maximum value $\rho_{\rm cr}$ (left panel in Fig.~\ref{fig4}). 
Therefore the collapse is halted, originating a bounce.

\begin{figure}
\begin{center}
\vspace{0.3cm}
\hspace{-0.8cm}
\includegraphics[type=pdf,ext=.pdf,read=.pdf,width=7.5cm]{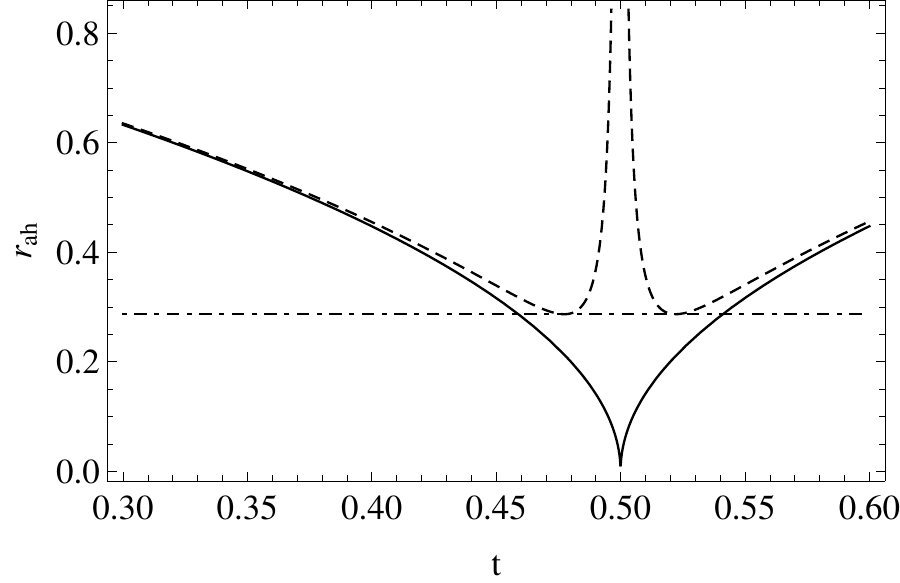}
\end{center}
\caption{The apparent horizon curve $r_{\rm ah}(t)$ for the classical radiation 
model (solid line) and the semiclassical model for $\gamma=1$ (dashed line). If 
the boundary of the cloud is taken smaller that $r_{\rm min}$ (dashed-dotted line) 
there are no trapped surfaces forming in the quantum inspired model. At the time 
of the bounce, when $\rho = \rho_{\rm cr}$ and gravity is turned off, the spacetime 
is flat and there is no horizon. This is a regime of asymptotic freedom.}
\label{fig6}
\end{figure}

Once again, an interesting question is what happens to the trapped
surfaces in this context. In the classical case, one must choose the 
boundary such that $r_{\rm b}<1/\sqrt{M_0}$ in order to avoid trapped 
surfaces at the initial time. In the semiclassical model, the condition becomes 
$r_{\rm b}<1/\sqrt{M_0(1-a_{\rm cr}^4)}$. The formation of trapped 
surfaces for the classical FRW radiation model is described by the time curve 
\begin{equation}
t_{\rm ah}(r)=t_{\rm s}-\frac{r^2\sqrt{M_0}}{2} \; ,
\end{equation}
 and the apparent horizon forms 
at the boundary of the cloud before the formation of the central singularity.
In our quantum inspired model, the apparent horizon curve is
\begin{equation}
r_{\rm ah}(t)=\frac{a^3}{\sqrt{M_0(a^4-a_{\rm cr}^4)}} \; .
\end{equation}
Like in the dust case, we can verify that there exists a minimum value $r_{\rm min}$, 
obtained as $t$ goes to 
\begin{equation}
t_{\rm min}=t_{\rm s}(\sqrt{1-a_{\rm cr}^4}-\sqrt{2}a_{\rm cr}^2) \; ,
\end{equation}
and it is given by
\begin{equation}
r_{\rm min}=r_{\rm ah}(t_{\rm min})=3^{3/4}\frac{a_{\rm cr}}{\sqrt{2M_0}} \; .
\end{equation} 
Therefore, if the boundary is taken so that $r_{\rm b}<r_{\rm min}$, no trapped 
surfaces form during the whole process of collapse and bounce (see Fig.~\ref{fig6}).
The existence of a minimum radius implies that there is a minimal mass $M_{\rm min}$ 
below which no apparent horizon can form.
In the general case, like for the dust model, an apparent horizon forms at
a time a little bit later than the classical prediction. It then disappears when the
density approaches the critical value $\rho_{\rm cr}$ and the gravitational force
is turned off ($r_{\rm ah}$ diverges). At the critical time, there is no horizon and the
spacetime is flat. After the bounce, the density decreases and a new apparent horizon 
forms. The latter eventually disappears when the density becomes too low.

\section{Concluding remarks}\label{conclusion}

Spacetime singularities, as obtained from exact solutions of Einstein's equations, 
are presumably the result of the breakdown of general relativity and they are 
supposed to be removed by quantum gravity corrections. So far, given the lack 
of a complete theory for quantum gravity, we do not really know how the issue of 
the formation of singularities is affected by quantum effects. It sounds plausible 
that singularities are bound to disappear once one treats the strong field regime 
within a suitable quantum gravitational framework. Toy models like the one 
discussed in the present paper may suggest possible scenarios. Classical singularities 
in general relativity can either be covered by a horizon or be naked. The issue 
of wether naked singularities can occur in a physically realistic scenario is still an 
open problem. Nevertheless, an analysis that takes into account quantum effects 
when the gravitational field becomes sufficiently strong not only affects the formation 
of the singularity, but it also has an impact on the structure of trapped surfaces.

The main result of our work is that in our model the outcome of the gravitational
collapse is not a black hole, in the sense of a region causally disconnected
from future null infinity. While we have not explicitly verified if outgoing null
geodesics launched from any point of the spacetime can propagate to null 
infinity, our results strongly suggest that this is indeed always the case. So,
there is no event horizon in these spacetimes and, in principle, 
the region where Planck scale effects become important could be visible to 
distant observers.

In our specific toy model with
$\gamma = 1$, we found that a homogeneous collapsing object reaches a critical
density. At this point, gravity is turned off and the apparent horizon disappears. After the
bounce, gravity becomes strong again and a new apparent horizon forms. The picture does
not seem to depend on the matter equation of state, and indeed we found the same
result for dust and radiation. The case $\gamma = 2$ is qualitatively similar
to the $\gamma = 1$ model for $t \le t_{\rm cr}$, with $t_{\rm cr} \rightarrow \infty$:
here there is no bounce and the asymptotic freedom regime where gravity becomes
weaker and weaker lasts for an infinite time, but gravity is never turned off, as
the critical density is reached only in an infinite time.

Classical singularities arising in astrophysical scenarios and not covered by any 
horizon suggest the possibility of observing regions where Planck scale physics 
produces detectable effects. This, in turn, may allow for the identification of a 
signature of quantum gravity and open the possibility of experimentally testing 
theories of quantum gravity via astrophysical observations. The long sought signature 
of quantum gravity, which has eluded any laboratory based hunt, might then be found 
in catastrophic astrophysical events.


\begin{acknowledgments}
This work was supported by the Thousand Young Talents 
Program and Fudan University.
\end{acknowledgments}



\begin{thebibliography}{99}

\bibitem{lqg1} 
  S.~Tsujikawa, P.~Singh and R.~Maartens,
  Class.\ Quant.\ Grav.\  {\bf 21}, 5767 (2004)
  [astro-ph/0311015];
  M.~Bojowald, G.~Calcagni and S.~Tsujikawa,
  Phys.\ Rev.\ Lett.\  {\bf 107}, 211302 (2011)
  [arXiv:1101.5391 [astro-ph.CO]];
  M.~Bojowald, G.~Calcagni and S.~Tsujikawa,
  JCAP {\bf 1111}, 046 (2011)
  [arXiv:1107.1540 [gr-qc]].
  
\bibitem{extra} 
  N.~Arkani-Hamed, S.~Dimopoulos and G.~R.~Dvali,
  Phys.\ Lett.\ B {\bf 429}, 263 (1998)
  [hep-ph/9803315];
  I.~Antoniadis, N.~Arkani-Hamed, S.~Dimopoulos and G.~R.~Dvali,
  Phys.\ Lett.\ B {\bf 436}, 257 (1998)
  [hep-ph/9804398].
  
\bibitem{grbs} 
  G.~Amelino-Camelia, J.~R.~Ellis, N.~E.~Mavromatos, D.~V.~Nanopoulos and S.~Sarkar,
  Nature {\bf 393}, 763 (1998)
  [astro-ph/9712103].
  
\bibitem{wccc} 
  R.~Penrose,
  Riv.\ Nuovo Cim.\  {\bf 1}, 252 (1969)
  [Gen.\ Rel.\ Grav.\  {\bf 34}, 1141 (2002)].
  
\bibitem{ns} 
  D.~M.~Eardley and L.~Smarr,
  Phys.\ Rev.\ D {\bf 19}, 2239 (1979);
  D.~Christodoulou,
  Commun.\ Math.\ Phys.\  {\bf 93}, 171 (1984);
  R.~P.~A.~C.~Newman,
  Class.\ Quant.\ Grav.\  {\bf 3}, 527 (1986);
  B.~Waugh and K.~Lake,
  Phys.\ Rev.\ D {\bf 38}, 1315 (1988);
  G.~Magli,
  Class.\ Quant.\ Grav.\  {\bf 14}, 1937 (1997);
  G.~Magli,
  Class.\ Quant.\ Grav.\  {\bf 15}, 3215 (1998);
  T.~Harada, K.~I.~Nakao and H.~Iguchi,
  Class.\ Quant.\ Grav.\  {\bf 16}, 2785 (1999)
  [gr-qc/9904073];
  T.~Harada and H.~Maeda,
  Phys.\ Rev.\ D {\bf 63}, 084022 (2001)
  [gr-qc/0101064];
  R.~Goswami and P.~S.~Joshi,
  Class.\ Quant.\ Grav.\  {\bf 19}, 5229 (2002)
  [gr-qc/0206086];
  R.~Giamb\`o, F.~Giannoni, G.~Magli and P.~Piccione,
  Commun.\ Math.\ Phys.\  {\bf 235}, 545 (2003)
  [gr-qc/0204030];
  R.~Goswami and P.~S.~Joshi,
  Phys.\ Rev.\ D {\bf 69}, 044002 (2004)
  [gr-qc/0212097];
  R.~Giamb\`o,
  J.\ Math.\ Phys.\  {\bf 47}, 022501 (2006)
  [gr-qc/0603120];
  P.~S.~Joshi, D.~Malafarina and R.~V.~Saraykar,
  Int.\ J.\ Mod.\ Phys.\ D {\bf 21}, 12500 (2012)
  [arXiv:1107.3749 [gr-qc]].

\bibitem{review}
  P.~S.~Joshi and D.~Malafarina,
  Int.\ J.\ Mod.\ Phys.\ D {\bf 20}, 2641 (2011)
  [arXiv:1201.3660 [gr-qc]].
  
\bibitem{zilong} 
  Z.~Li and C.~Bambi,
ÊÊPhys.\  Rev.\ D {\bf 87}, 124022 (2013)
ÊÊ[arXiv:1304.6592 [gr-qc]]. 
  
\bibitem{tests} 
  C.~Bambi and K.~Freese,
  Phys.\ Rev.\ D {\bf 79}, 043002 (2009)
  [arXiv:0812.1328 [astro-ph]];
  C.~Bambi, K.~Freese, T.~Harada, R.~Takahashi and N.~Yoshida,
  Phys.\ Rev.\ D {\bf 80}, 104023 (2009)
  [arXiv:0910.1634 [gr-qc]];  
  C.~Bambi, T.~Harada, R.~Takahashi and N.~Yoshida,
  Phys.\ Rev.\ D {\bf 81}, 104004 (2010)
  [arXiv:1003.4821 [gr-qc]];
  C.~Bambi and N.~Yoshida,
  Class.\ Quant.\ Grav.\  {\bf 27}, 205006 (2010)
  [arXiv:1004.3149 [gr-qc]];
  C.~Bambi,
  Europhys.\ Lett.\  {\bf 94}, 50002 (2011)
  [arXiv:1101.1364 [gr-qc]];
  C.~Bambi,
  JCAP {\bf 1105}, 009 (2011)
  [arXiv:1103.5135 [gr-qc]];
  C.~Bambi, F.~Caravelli and L.~Modesto,
  Phys.\ Lett.\ B {\bf 711}, 10 (2012)
  [arXiv:1110.2768 [gr-qc]];
  Z.~Li and C.~Bambi,
  JCAP {\bf 1303}, 031 (2013)
  [arXiv:1212.5848 [gr-qc]]:
  C.~Bambi and D.~Malafarina,
ÊÊarXiv:1307.2106 [gr-qc];
  A.~N.~Chowdhury, M.~Patil, D.~Malafarina and P.~S.~Joshi,
  Phys. Rev. D {\bf 85}, 104031 (2012) 
  [arXiv:1112.2522 [gr-qc]];
  P.~S.~Joshi, D.~Malafarina and R.~Narayan
  Class. Quantum Grav. {\bf 28}, 235018 (2011)
  [arXiv:1106.5438 [gr-qc]];
  D.~Pugliese, H.~Quevedo and R.~Ruffini R, 
  Phys. Rev. D {\bf 84} 044030 (2011)
  [arXiv:1105.2959 [gr-qc]]; 
  D.~Pugliese, H.~Quevedo and R.~Ruffini, 
  Phys. Rev. D {\bf 83}, 104052 (2011)
  [arXiv:1103.1807 [gr-qc]]; 
  Z.~Kovacs and T.~Harko, 
  Phys. Rev. D {\bf 82}, 124047 (2010)
  [arXiv:1011.4127 [gr-qc]];
  K.~S.~Virbhadra and G.~F.~R.~Ellis, 
  Phys. Rev. D {\bf 65}, 10300 (2002).
  
\bibitem{Mode}
  L.~Modesto, J.~W.~Moffat and P.~Nicolini,
  Phys.\ Lett.\ B {\bf 695}, 397 (2011)
  [arXiv:1010.0680 [gr-qc]];
  L.~Modesto,
  Phys.\ Rev.\ D {\bf 86}, 044005 (2012)
  [arXiv:1107.2403 [hep-th]];
  L.~Modesto,
  Astron.\ Rev.\  (in press)
  [arXiv:1202.3151 [hep-th]].
  
\bibitem{Nico}
  L.~Modesto,
  Int.\ J.\ Theor.\ Phys.\  {\bf 45}, 2235 (2006)
  [gr-qc/0411032];
  L.~Modesto,
  Class.\ Quant.\ Grav.\  {\bf 23}, 5587 (2006)
  [gr-qc/0509078];
  P.~Nicolini, A.~Smailagic and E.~Spallucci,
  Phys.\ Lett.\ B {\bf 632}, 547 (2006)
  [gr-qc/0510112];
  L.~Modesto,
  Int.\ J.\ Theor.\ Phys.\  {\bf 47}, 357 (2008)
  [gr-qc/0610074];
  L.~Modesto,
  Adv.\ High Energy Phys.\  {\bf 2008}, 459290 (2008)
  [gr-qc/0611043];
  P.~Nicolini,
  Int.\ J.\ Mod.\ Phys.\ A {\bf 24}, 1229 (2009)
  [arXiv:0807.1939 [hep-th]];
  L.~Modesto and I.~Premont-Schwarz,
  Phys.\ Rev.\ D {\bf 80}, 064041 (2009)
  [arXiv:0905.3170 [hep-th]];
  S.~Hossenfelder, L.~Modesto and I.~Premont-Schwarz,
  Phys.\ Rev.\ D {\bf 81}, 044036 (2010)
  [arXiv:0912.1823 [gr-qc]];
  L.~Modesto,
  Int.\ J.\ Theor.\ Phys.\  {\bf 49}, 1649 (2010);
  L.~Modesto and P.~Nicolini,
  Phys.\ Rev.\ D {\bf 82}, 104035 (2010)
  [arXiv:1005.5605 [gr-qc]].

\bibitem{cosmology} 
  M.~Bojowald,
  Phys.\ Rev.\ Lett.\  {\bf 86}, 5227 (2001)
  [gr-qc/0102069];  
  A.~Ashtekar, T.~Pawlowski and P.~Singh,
  Phys.\ Rev.\ D {\bf 73}, 124038 (2006)
  [gr-qc/0604013];
    A.~Ashtekar, T.~Pawlowski and P.~Singh,
  Phys.\ Rev.\ D {\bf 74}, 084003 (2006)
  [gr-qc/0607039];
  E.~Wilson-Ewing,
  JCAP {\bf 1303}, 026 (2013)
  arXiv:1211.6269 [gr-qc].
  
\bibitem{rbh} 
  M.~Bojowald, R.~Goswami, R.~Maartens and P.~Singh,
  Phys.\ Rev.\ Lett.\  {\bf 95}, 091302 (2005)
  [gr-qc/0503041];
  R.~Goswami, P.~S.~Joshi and P.~Singh,
  Phys.\ Rev.\ Lett.\  {\bf 96}, 031302 (2006)
  [gr-qc/0506129];
  R.~Casadio, S.~D.~H.~Hsu and B.~Mirza,
  Phys.\ Lett.\ B {\bf 695}, 317 (2011)
  [arXiv:1008.2768 [gr-qc]].
  
\bibitem{tavakoli} 
  Y.~Tavakoli, J.~Marto and A.~Dapor,
  arXiv:1303.6157 [gr-qc]. 
  
\bibitem{os} 
  J.~R.~Oppenheimer and H.~Snyder,
  Phys.\ Rev.\  {\bf 56}, 455 (1939).  

\bibitem{visser} C. Barcelo and M. Visser, 
 Int. J. Mod. Phys. D {\bf 11}, 1553 (2002)
 [arXiv:gr-qc/0205066 [gr-qc]];
 P. Martin-Moruno and M. Visser, arXiv:1305.1993  [gr-qc].
  
\bibitem{ltb} 
  G.~Lemaitre,
  Annales Soc.\ Sci.\ Brux.\ Ser.\ I Sci.\ Math.\ Astron.\ Phys.\ A {\bf 53}, 51 (1933)
  [Gen.\ Rel.\ Grav.\  {\bf 29}, 641 (1997)];
  R.~C.~Tolman,
  Proc.\ Nat.\ Acad.\ Sci.\  {\bf 20}, 169 (1934)
  [Gen.\ Rel.\ Grav.\  {\bf 29}, 935 (1997)];
  H.~Bondi,
  Mon.\ Not.\ Roy.\ Astron.\ Soc.\  {\bf 107}, 410 (1947).

\bibitem{matching} 
  W.~Israel,
  Nuovo Cim.\ B {\bf 44S10}, 1 (1966)
  [Erratum-ibid.\ B {\bf 48}, 463 (1967)]
  [Nuovo Cim.\ B {\bf 44}, 1 (1966)].
  P.~S.~Joshi and I.~H.~Dwivedi,
  Class.\ Quant.\ Grav.\  {\bf 16}, 41 (1999)
  [gr-qc/9804075];
  R.~Giamb\`o,
  Class.\ Quant.\ Grav.\  {\bf 22}, 2295 (2005)
  [gr-qc/0501013].

\bibitem{dust} 
  P.~S.~Joshi and I.~H.~Dwivedi,
  Phys.\ Rev.\ D {\bf 47}, 5357 (1993)
  [gr-qc/9303037];
  P.~S.~Joshi, N.~Dadhich and R.~Maartens,
  Phys.\ Rev.\ D {\bf 65}, 101501 (2002)
  [gr-qc/0109051].

\bibitem{santos} 
  N.~O.~Santos,
  Mon.\ Not.\ Roy.\ Astron.\ Soc.\  {\bf 216}, 403 (1985).

\bibitem{pjm}  
  R.~Goswami, P.~S.~Joshi, C.~Vaz and L.~Witten,
  Phys.\ Rev.\ D {\bf 70}, 084038 (2004) 
  [arXiv:gr-qc/0410041 [gr-qc]];
  M.~Patil, P.~S.~Joshi and D.~Malafarina,
  Phys.\ Rev.\ D {\bf 83}, 064007 (2011)
  [arXiv:1102.2030 [gr-qc]].

\end{thebibliography}
\end{document}